\documentclass{elsart}

\usepackage{epsfig}

\usepackage{amssymb}

\usepackage{color}

\begin{document}

\begin{frontmatter}



\title{The Diffuse Interstellar Bands: A Major Problem in Astronomical
Spectroscopy}

\author{Peter J. Sarre}

\address{School of Chemistry, The University of Nottingham, University Park,
Nottingham, NG7 2RD, United Kingdom}

\begin{abstract}
A critical review of the very long-standing problem of the diffuse
interstellar bands is presented with emphasis on spectroscopic
aspects of observational, modelling and laboratory-based research.
Some research themes and ideas that could be explored theoretically
and experimentally are discussed. The article is based on the
Journal of Molecular Spectroscopy Review Lecture presented at the
60th Ohio State University International Symposium on Molecular
Spectroscopy, June 2005.

% Text of abstract
\end{abstract}

%%%%%%%%%%%%%%%%%%%%%%%%%%%%%%%%%%%%%%%%%%%%%%%%%%%%%%%%%%%%%%%%%%%%%%%
\end{frontmatter}
%%%%%%%%%%%%%%%%%%%%%%%%%%%%%%%%%%%%%%%%%%%%%%%%%%%%%%%%%%%%%%%%%%%%%%

% main text
\section{Introduction and a brief history}
%\label{}
The longest standing challenge in astronomical spectroscopy is to
assign the diffuse interstellar bands. These absorption features
appear in spectra recorded towards stars that lie behind diffuse
interstellar clouds and fall largely in the visible part of the
electromagnetic spectrum. The first observational record is that
made by Mary Lea Heger during her PhD research at Lick Observatory
\cite{heg22} but systematic studies began with the work of Paul
Merrill \cite{mer34,mer36}. In 1934 he wrote \cite{mer34}: `Recent
observations at Mount Wilson, made chiefly to investigate the
interstellar sodium lines, have disclosed four additional detached
lines in the yellow and red whose approximate wave-lengths are
5780.4, 5796.9, 6283.9, and 6613.9 \AA, respectively. These lines,
found in types Oa to A4, behave like interstellar lines with regard
to occurrence, intensity, and displacement.  Instead of being narrow
and sharp, however, as interstellar lines should be, they are
somewhat widened and have rather diffuse edges.  Their chemical
identifications have not been found.  The widening of the lines and
the difficulty of identification make the problem of their atomic
origin an interesting one.' Merrill's studies predated by a few
years the discovery of the first interstellar molecules, namely CH,
CH$^{+}$ and CN, that was made possible through a combination of
observational and laboratory spectroscopy as described by Herzberg
\cite{herz88}.

The early diffuse band researchers would surely not have thought
that the solution to the diffuse band problem would still be so
elusive. It is a particularly remarkable situation given the huge
scientific and technological advances made over many decades. This
includes the development of radio and infrared astronomy that has
allowed the detection of over one hundred molecules in interstellar
clouds. However, the spectral characteristics of the bands are now
much better defined and this has aided progress, most notably
through stimulating laboratory research and in ruling out various
suggested potential carriers. A major difficulty remains in deciding
which laboratory and theoretical experiments should be performed and
this is where further selected observations and spectral modelling
can contribute.

A major article was published by George Herbig in 1975 and includes
figure 1. which illustrates the principal characteristics of the
interstellar (IS) bands \cite{her75}.  Spectra recorded towards the
star HD 183143 which is reddened by intervening dust and an
unreddened reference standard are shown with some of the key bands
indicated, \emph{viz.} $\lambda\lambda$5780, 5797 and 5849 (upper),
together with $\lambda$6195 and $\lambda$6177 (lower).  The bands
widths differ in figure 1, showing narrow ($\lambda$6195),
intermediate ($\lambda$5780) and broad ($\lambda$6177) examples.
They are described as being `diffuse' as their widths are greater
than those of absorption features of known atoms and molecules along
the line of sight. Described sometimes as fine structure on the
interstellar extinction curve, this aspect is illustrated in figure
2, again for HD 183143, which appeared in a second key review
published by Herbig twenty years later \cite{her95}.  A number of
other articles have been written covering diffuse band research
\cite{smi77,bro87,sno95,wil96,sno01,kre02} as well as a new general
review of diffuse cloud chemistry and spectroscopy \cite{sno06}.
Data for a substantial part of the spectrum are summarized in figure
3.

In this paper the very broad 2175 \AA\ absorption feature, which may
be related to the diffuse band problem, is not discussed. Also only
brief reference is made to the `unidentified' infrared bands though
their carriers are quite possibly related to the diffuse band issue.
I concentrate on the problem as viewed from a spectroscopic
perspective, describe recent results, and suggest some new areas
that may help in the search for an assignment.

\section{Characteristics of the diffuse interstellar band spectrum}

Over three hundred \emph{diffuse} interstellar bands have been
documented, the most striking characteristic being their widths
which range between $\sim$2 and $\sim$100~cm$^{-1}$ \cite{her95}.
Their diffuseness is most commonly attributed to short lifetimes of
the excited states of the transitions as discussed by Smith et al.
\cite{smi77}. That the bands arise in \emph{interstellar} rather
than stellar or circumstellar material is beyond doubt at least for
those cases where this has been investigated. In a type of study
undertaken by Merrill \cite{mer36}, observations towards the binary
HD 23180 on successive observing nights (figure 4.) show that the
diffuse bands, together with interstellar sodium, are `stationary'
and do not share the Doppler shift of stellar features such as the
$\lambda$5875 line of photospheric Helium \cite{kre93}. The word
`\emph{band}' is commonly used, although with the exception of some
data from high-resolution studies there is little evidence that the
widths are due to the unresolved blending of individual rotational
lines as could arise in the electronic spectrum of a gas-phase
molecule.  The spectra are confined to the range between $\sim$4000
and ~$\sim$13,000 \AA\ which is roughly equivalent to photon
energies in the 1-3 eV range. The carriers are ubiquitous, being
found in numerous reddened Galactic lines of sight and also in
external galaxies including the Magellanic clouds
\cite{mor87,vla87,ehr02,cox06}, starburst galaxies
 \cite{hec00}, NGC 1448 \cite{sol05} and a damped Ly$\alpha$
 system at $z_{a}$ = 0.524 \cite{jun04}.

There is a good correlation between the band strengths and the
reddening index $E_{B-V}$. This indicates that the dust particles
that cause optical extinction are probably associated with the
diffuse band carriers although these relatively large micron-sized
grains are very unlikely to be directly responsible for the diffuse
band absorptions. The extent of the correlation for $\lambda$5780 is
illustrated in figure 5, \cite{kre99} which also highlights the real
scatter that exists, the origin of which is not understood. Examples
where this general relationship with extinction does not appear to
hold well are rare but significant \cite{sno02}. A second important
correlation is that with the column density of hydrogen atoms rather
than with molecular hydrogen \cite{her93}

Key spectroscopic characteristics include a lack of regularity in
the wavenumbers of the bands, as might be expected to be present in
the electronic spectrum of a molecule, and a lack of common band
widths as would occur if the same excited state with a short
lifetime were accessed by more than one transition.  However, if all
optical excitations originate in the same quantum level on account
of the low temperature of the cloud, then appearance of common
widths is unlikely. The breadth of the bands has long been
recognised as an extraordinary property which is considered by many
to be a defining characteristic and one that might be expected to
play a major role in finding a solution. Given the generally good
correlation with $E_{B-V}$, this led to proposals based on optical
absorptions of carriers that are in or on grains.  In this case the
widths are generally attributed to broadening of a transition due to
the condensed phase environment. However, the alternative proposal
that the bands originate from electronic transitions in gas-phase
molecules has gained ground particularly over the last \emph{c.} 25
years.  A key aspect is the recognition that large molecules can
accommodate photon absorption without destruction and give rise to
broad spectra due to fast internal relaxation processes
\cite{smi77}.

There is a wide range of data available from both photographic
\cite{her75,sno77} and CCD surveys with those up to the year 2000
described by Tuairisg et al. \cite{tua00}, to which further studies
have been added \cite{wes00,mcc02}.

\section{Proposals for the origin of the diffuse bands}
The idea of molecules being responsible for the bands stretches back
to the 1930s, followed by a period in which a dust grain origin was
preferred.  A molecular origin was championed for many years by
Herzberg, although the specific idea of predissocation or
preionisation being responsible for the widths is not now favoured.
Suggestions for assignments include colour centres, lattice defects,
the hydrogen anion, porphyrins, carbon chains, charge transfer
transitions and molecular hydrogen. To this list of possibilities
polycyclic aromatic hydrocarbons, simple or encapsulated fullerenes
and carbon nanotubes may be added. It is of interest that the
discovery of C$_{60}$ was made in experiments motivated in part by
the search for a solution to the diffuse band problem. The full
history has been reviewed in a number of papers
\cite{her75,her95,bro87,sno95,sno01}.

The discovery of widespread `unidentified' infrared (UIR) emission
bands arising from polycyclic aromatic hydrocarbons (PAHs) in
nebular and other regions (and also recently in the general
interstellar medium) has led to the development of the `PAH
hypothesis' \cite{leg85,van85} and a huge laboratory and theoretical
effort focussed on PAHs as possible carriers
\cite{sal96,rui02,wei03,hal05}.

\section{Spectroscopic aspects: the molecular hypothesis}

This section is written under the assumption that the diffuse bands
are caused by electronic transitions of gas-phase molecules which
under interstellar conditions have low internal temperatures. It may
be assumed that all molecules will be in their lowest vibrational
state, with the rotational temperature ranging between about 3 K for
a polar molecule as found for CN and \emph{c.} 100 K as for C$_{2}$
in some lines of sight. The shapes and strengths of the diffuse
bands are many and varied, ranging from narrow bands with asymmetry
and substructure such as $\lambda\lambda$5797, 6614 and 6376/6379,
to the broadest and strongest $\lambda$4428 (formerly $\lambda$4430)
band which exhibits no fine structure even observed with 10$^6$
resolving power \cite{sno02a}; it obeys a perfect fit to a
Lorentzian function \cite{sno02b}.

\subsection{Regularities and correlations}

Given the importance of pattern recognition in spectroscopic
assignments, attempts have been made to find links between bands.
Herbig \cite{her88} noted a small set of bands near 6800 \AA\ with
intervals of $\sim$35 cm$^{-1}$ and possible assignments in terms of
C$_n$H$_2$-type polyene molecules have been offered
\cite{gli95,sch00}. The core idea is that the bands are $K$
sub-bands and the separations of $\sim$35~cm$^{-1}$ correspond to
$\sim$4$A$ where $A$ is the rotational constant associated with the
$a$ inertial axis \cite{gli95,sch00}. However, only those sub-bands
with a statistical weight of 3 are clearly observed so there remains
some doubt as to whether this is the correct explanation
\cite{tad05}. A set of possibly linked bands has been put forward by
Weselak et al. \cite{wes01} who tentatively suggested that
$\lambda\lambda$5809, 6196, 6614 and 6660 might be from the same
carrier but this awaits further study.

Numerous investigations of possible correlations between the
strengths of various bands have been made, resulting in the concept
of `families' first described by Kre{\l}owski and Walker in 1987
\cite{kre87} and since developed by others (see \cite{cox05} for a
summary). One case where there is a particularly good correlation is
that between $\lambda$6614 and $\lambda$6196 which has been found to
have a correlation coefficient in equivalent width of 0.98$\pm$0.18
over 62 lines-of-sight \cite{mou99}. Following a detailed
statistical assessment a value of 0.985 was reported at the
Symposium based on APO survey results \cite{mcc05}. Even in this
case the question as to whether the two bands arise from the same
carrier remains open; from high resolution and high signal-to-noise
spectra Galazutdinov et al. have drawn the conclusion that the ratio
of equivalent widths of the two diffuse bands is not always exactly
the same \cite{gal02}.

A number of weak diffuse bands appear to be correlated with the
presence of C$_2$ \cite{tho03} which suggests that the formation
chemistry for these band carriers may be similar to that of C$_2$
(and C$_3$). Some of the `C$_2$' bands occur in pairs with very
similar separations of 20.7, 20.7 and 20.9 cm$^{-1}$ which may arise
from spin-orbit interaction. A further study of these bands has been
made from which it is deduced that the `C$_2$ diffuse band'
designation is premature and that it is unlikely that the bands are
related to C$_2$ \cite{gal06}. In the event that any two diffuse
bands originate from the same level of the same carrier then the
correlation must of course be perfect.  However, in seeking insight
into the problem, there is considerable merit in finding families of
bands that generally behave in a similar way.

\subsection{High-resolution spectra and modelling}

Interest in high-resolution spectroscopy has grown since the
discovery of fine structure in some of the narrower diffuse bands
such as $\lambda$6614 \cite{sar95,ehr96}. The $\lambda$6614 band has
been found to have three main components that have the appearance of
unresolved P, Q and R branches of an electronic transition of a
large molecule \cite{sar95}. Building on the principles first
applied to diffuse band profiles by Danks and Lambert \cite{dan76},
the profile of $\lambda$6614 was fitted in a $\chi$-squared
minimisation procedure to derive molecular constants and rotational
temperatures  as shown in figure 6 \cite{ker96}. A range of
molecular geometries and types of electronic transition was
explored, the best fit to the profile being achieved for an oblate
symmetric top like coronene where the $\tilde{A}$$^1$B$_2$ -
$\tilde{X}$$^1$A$_1$ transition is vibronically induced as the
origin band is symmetry forbidden. A plausible fit was obtained for
a range of low rotational temperatures and it was noted that a
C$_{18}$ (monocyclic cumulenic) ring could also be a candidate
\cite{ker96}. Neither of these specific molecules is a diffuse band
carrier as the strongest band of coronene at 4031.4 \AA\ does not
appear in the spectra of diffuse clouds \cite{sar98} and the
$\tilde{A}$$^1$A$_2$$^{\prime\prime}$ -
$\tilde{X}$$^1$A$_1$$^{\prime}$ transition of C$_{18}$ occurs at
5928.5 \AA\ which is also not detected \cite{mai06}. However, this
modelling study did illustrate the principle that the fine structure
could be reproduced in terms a molecular electronic transition and
at typical interstellar temperatures. An ultra-high resolution
spectrum of $\lambda$5797 has revealed an even higher level of fine
structure \cite{ker98} but this has not yet been modelled
successfully. A rough estimate of the size of carrier can be made by
taking the separation of the `P' and `R' branch peaks. Using this
approach Ehrenfreund and Foing inferred that the spectra of
$\lambda\lambda$6614, 5797 and 6379 were consistent with PAH
molecules with more than 40 carbon atoms, chains of $\sim$12-18
carbon atoms, rings with 30 carbons or fullerenes \cite{ehr96}.
Schulz et al. have suggested that a polyene with 9-13 carbon atoms
could account for the $\lambda$6614 fine structure \cite{sch00}. All
of these approaches assume thermally populated rotational levels. A
more detailed treatment of level populations has been presented by
Mulas \cite{mul98}. In a study of the $\lambda\lambda$6196 and 6379
bands, it was reported that rotational contour models did not fit
either profile well, there being more structure in the model than is
observed \cite{wal01}. The level of agreement was improved by
invoking a broadening of $\sim$0.2-0.3 cm$^{-1}$ \cite{wal01}. In
related work, contour modelling predictions have been pursued for
PAH-type systems and fullerenes \cite{cos90,edw93} and spectra of
the $\lambda$5850 band and nearby features have been discussed in
terms of possible rotational branch structure \cite{jen96}.

A growing area of observational research is the study of profile
variation that might be indicative of changes in level populations,
physical state or local interstellar conditions. Neglecting band
broadening that occurs simply because there are two or more clouds
in a line-of-sight \emph{e.g.} HD 183143 \cite{her82}, there are
reports of unusual band widths and shifts in spectra taken towards
Orion stars \cite{por92,kre99a} and HD 34078 \cite{gal06a}.  The
origin of this effect is not established.

\subsection{Carbon isotope structure}
A completely different hypothesis for the origin of the `triplet'
structure of the $\lambda$6614 band has been put forward by Webster
\cite{web96}. It is suggested that each of the component peaks
arises from a separate $^{13}$C isotopic modification of the same
molecule where the relative positions of the components are
determined by a vibrational isotope shift in the excited electronic
state. The work highlights the importance of considering the
$^{13}$C isotope, provides a good fit to the spectrum and has also
been applied to the $\lambda$6196 profile in connection with the
zero-point vibrational isotope shift \cite{web04}. Further
observations of $\lambda\lambda$5797 and 6614 have been taken as
evidence in support of carbon isotope structure in the spectrum
\cite{wal00}. However, in a recent study of $\lambda$6614 at high
resolution, it is reported that the positions (wavelengths) of two
of the substructure peaks vary according to the line-of-sight
\cite{cam04}. While the wavelength of the strongest central `Q'
component does not change, the `Q-R' and `Q-P' separations increase
and decrease in concert though to different extents. This is
reminiscent of the behaviour of molecular band contours as a
function of temperature and a range of $\sim$21-25 K for $T_{rot}$
was deduced. From these observations it was concluded that the
observed variations rule out the idea that the $\lambda$6614
substructure arises from isotope shifts \cite{cam04}.

\section{Circumstellar and nebular environments}

Study of the diffuse band carriers in localised astrophysical
environments such as stars and nebulae offers the potential to
investigate their spectroscopic transitions under different
excitation conditions and potentially to learn something of their
formation mechanism. Unfortunately such data are extremely rare in
spite of many observational searches.  The circumstellar shell of
the nearby carbon star IRC +10$^o$216 and the Red Rectangle nebula
are of interest in this context.

\subsection{The circumstellar shell of IRC +10$^o$216}

One of the more likely environments in which the diffuse band
carriers might be found is in the circumstellar envelope of a
mass-losing carbon star. A summary of observations made in the
search for circumstellar diffuse band absorptions in carbon stars
and planetary nebulae is given in references
\cite{zac99,zac03,mau04} but with very largely negative or uncertain
results. The most studied circumstellar shell in astrochemistry is
IRC +10$^o$216 in which numerous molecules have been detected by
radio, infrared and optical spectroscopy.  A search for diffuse band
absorption due to carriers in the envelope was made by recording
spectra towards a star (Star 6) which lies behind part of the
extended shell as shown in figure 7 \cite{ken02}.  No absorption at
the expected position for $\lambda$6614 was found. Although the
 precursors to the diffuse band carriers may well be formed
in such objects, it seems that further chemical or UV processing is
required before they become spectroscopically active in the
interstellar medium.

\subsection{The Red Rectangle}

The importance of the Red Rectangle nebula and its central binary
star HD 44179 was first signalled in a paper by Cohen et al. in 1975
\cite{coh75}. This biconical nebula displays many unusual unassigned
spectroscopic features from the infrared to the far-UV including
strong optical emission bands \cite{war81} with wavelengths which
lie close to some of the diffuse bands seen in absorption
\cite{sar91,fos91,sca92,sar95a,van02}. This is illustrated in figure
8 for a small part of the region near 5800~\AA\ where the Red
Rectangle spectrum at an offset from the central star of
7.5$\pm$2.5$''$ is compared with the diffuse band absorption
spectrum towards HD 183143.  Measurements of the peak wavelengths
and widths of the strongest emission bands show that they change
with offset from the exciting star and that these two attributes
converge \emph{towards} the values seen in diffuse band absorption.
However, even at the highest offset so far attained they are not
coincident with the interstellar absorption data
\cite{sar95a,van02,gli02,sha06}. A possible explanation is that the
internal temperature of the carrier is `locked' as occurs for C$_2$.
Resonance fluorescence of the Swan 5165~\AA\ (0,0) band of C$_2$
(d$^3$$\Pi$$_g$ - a$^3$$\Pi$$_u$) is detectable out to at least
13$''$ offset in the nebula.  This means that the metastable
a$^3$$\Pi$$_u$ state with $\sim$~700 cm$^{-1}$ excitation is
significantly populated.  Moreover the high-offset C$_2$ spectrum
has a band head which indicates a rotational temperature in excess
of that found \emph{via} the C$_2$ Phillips A$^1$$\Pi$$_u$
-X$^1$$\Sigma$$_g^+$ system (only) in diffuse interstellar clouds.

The shapes of two of the most prominent emission bands near 5800 and
6615 \AA\ are reminiscent of molecular rotational contours
\cite{sca92,sar95a} but there may also be contributions from
unresolved vibrational sequence band structure \cite{sha06}. The
bands also appear weakly in the spectrum of the R CrB star V854~Cen
at minimum light \cite{rao93}. Whether the set of unidentified Red
Rectangle emission bands and diffuse bands arise from the same
carriers as suggested \cite{sar95a} will probably be proven
definitively only when the carriers are identified and the spectra
are reproduced in the laboratory. However the likelihood of this
being the case is sufficient to have stimulated new laboratory
searches for `diffuse band' spectra in emission, but so far without
success.

\section{Spatial distribution of the diffuse band carriers}

Considerable research effort has been committed to searching for
correlations between bands by recording spectra along many lines of
sight through different types of clouds.  In a complementary
approach we have commenced a new observing program in which spectra
are recorded at extremely high signal-to-noise ratio towards binary
stars that are separated in light path by only $\sim$100-10,000 AU.
This simplifies the situation because, in contrast to widely
separated unrelated lines of sight, the cloud conditions for the two
lines-of-sight should have some chemical and physical similarities.
However, from an observational viewpoint, discerning small
differences in diffuse band strengths is very challenging and hence
requires long integration times.  The work is motivated in part by
the subject of this article, but in addition there is much general
astronomical interest in determining the level of `small-scale
structure' of the interstellar medium.

Spectra recorded using UCLES at the Anglo-Australian Telescope
towards $\rho$ Oph A and $\rho$ Oph B are shown in figure 9. These
stars are separated by $3.1''$ which at a distance of ~$\sim$120 pc
results in a sky-projected separation of $\sim$370 AU.  Taking the
\emph{ratio} of the spectra towards $\rho$ Oph A and B yields the
lower trace of figure 9. Equivalent width measurements give quite a
large increase in strength of both $\lambda$5780 and $\lambda$5797
of \emph{c.} 5\% from A to B. Initial results indicate that this
type of study will produce further information on diffuse band
families, relationships between diffuse band carriers and
atoms/molecules along the same lines-of-sight, and on the physical
conditions in the cloud.

\section{Solving the problem}

Most researchers consider that the diffuse bands arise from a set of
carbon-based absorbers with size between the dust grains that cause
 visual extinction and the largest carbon molecule so far
identified optically in diffuse clouds, C$_{3}$ \cite{haf95,mai01}.
In this section a brief overview of laboratory research is given
followed by a few speculative reflections on the diffuse band
problem.  Attention is drawn to some areas that deserve further
study including clues that may be available from other areas such as
the UIR bands and the Red Rectangle.

%The next molecules in the neutral carbon chain series C$_{4}$ and C$_{5}$, have not been detected \cite{mai02,gal02a}.

\subsection{Laboratory experiments}
An enormous wealth of laboratory electronic spectra of carbon
chains, PAHs and fullerenes in neutral and ionised forms has been
recorded in the gaseous phase and in matrices at low temperature.
Although not always yielding transition frequencies as precise as
desirable due to uncertainty in the degree of internal excitation in
the gas phase and the influence of the matrix in condensed phase
work, there is little doubt in the author's view that a
correspondence with any of the most prominent features in the
diffuse band spectrum would be apparent. In some instances there are
also uncertainties in the astrophysical data due to stellar spectral
contamination and poorly defined velocity distributions in the
cloud, coupled with the problem of the unknown carrier transition
rest frequency. No attempt to describe this work in full is made
here; comparisons between laboratory and astrophysical data can be
found in \cite{sal99,ful00,mot00}. Recent `near-misses' or cases
where the data do not allow a definitive conclusion to be made
include studies of C$_{14}$H, C$_{7}^{-}$, C$_{3}$H$_{2}^{-}$,
C$_{10}$H$_{8}^{+}$ and C$_{5}$H$_{5}$.  The intriguing possibility
of C$_{60}^{+}$ being responsible for two red bands awaits gas-phase
spectra \cite{gal00}. There is also a possible correspondence
between the gas-phase origin band of CH$_{2}$CN$^{-}$ and a weak
diffuse band at 8037 \AA\ \cite{sar00} but this needs higher quality
observational spectra to determine whether predicted satellite $K$
sub-bands are present. This partial list of recently studied spectra
illustrates the surge in laboratory effort stimulated by the diffuse
band problem that has led to important advances in the spectroscopy
of these molecules through often very challenging experiments. There
do remain some areas that are virtually untouched including the
laboratory electronic spectra of gas-phase carbon polyene chains,
protonated, anionic and heteroatom-containing PAHs, fullerenes and
carbon nanotubes.

\subsection{Heteroatoms in and on PAHs}

Almost all astrophysically oriented research on the electronic
transitions of neutral and ionised PAHs considers them to contain
carbon and hydrogen only. However, the rich field of colour
chemistry contains hundreds of polyaromatic compounds in which
nitrogen and oxygen atoms are present either on the periphery or as
part of the aromatic ring system. Their inclusion is crucial to the
very strong absorption and emission of dye-type molecules in the
visible part of the spectrum. Both types of chemical modification of
PAHs by heteroatoms are found in meteorites. To take an example,
proflavine absorbs strongly in the blue region at 445 nm (in water
at pH 7) with a molar extinction coefficient of \emph{c.} 40,000.
Some plausible diffuse band carrier candidates do not require
heteroatom substitution in the ring system. An example is the
quinones which contain a C=O group. In the Red Rectangle nebula,
which has a mixed carbon-oxygen chemistry, a rare infrared emission
band at 6.0 $\mu$m is likely due to the C=O stretch of an oxygenated
PAH \cite{pee02}. While terrestrial dye-type molecules are
relatively small, the principle should carry over to larger PAH
systems that are likely to be present in the interstellar medium and
therefore warrant laboratory and theoretical study. Some aspects of
the electronic transitions of PAHs with exo- and endohedral nitrogen
have recently been considered in connection with possible diffuse
band carriers \cite{hud05}.

\subsection{Clues from the Red Rectangle?}

The Red Rectangle displays the UIR bands strongly from both the
central star and the nebula and this has led to the idea that PAHs
are responsible for the unidentified optical emission bands
including those close in wavelength to some diffuse bands. While
there is no correlation between the 3.3 $\mu$m `PAH' emission
feature and the unidentified optical red/yellow emission bands
\cite{ker99}, blue fluorescence has been discovered \cite{vij04}
that does follow the 3.3 $\mu$m IR emission \cite{vij05}. This, with
other factors, has led to the proposed presence of small gas-phase
PAHs in the Red Rectangle as they are known to fluoresce in the
near-UV \cite{vij04,vij05}. However, recent results from the Spitzer
satellite have shown that the carbon-oxygen chemistry in the Red
Rectangle is mixed \cite{mar05} with oxygen-rich material probably
being present in the nebula.  Hence inorganic or organometallic
molecules (or dust grains) as carriers of the unidentified optical
electronic emission spectra (such as in figure 8) and the `blue
fluorescence' \cite{vij04,vij05} cannot be ruled out. It is notable
that a set of natural olivine samples demonstrate UV-pumped
photoluminescence of hole centres in the same `blue fluorescence'
range \cite{bak95} as that attributed \cite{vij05} to small PAHs in
the Red Rectangle. It is potentially of interest that the 5800 \AA\
bands fall very close to the `yellow' bands of gas phase FeO. While
the observed spectrum does not arise from the gas-phase iron oxide
molecule, perhaps this is more than a coincidence.

% \appendix

% \section{}
% \label{}

Acknowledgements

That part of the work described which has been undertaken at
Nottingham has been conducted in collaboration with other
researchers to whom I am greatly indebted. I wish particularly to
thank Paul Boichat, Martin Cordiner, Stephen Fossey, Rob Hibbins,
Mark Hurst, Tim Kendall, Tom Kerr, June McCombie, Janet Miles, Jules
Russell, Arfon Smith and Radmila Topalovic. The observational work
described was awarded by the UK Panel for the allocation of
telescope time (PATT) and the research made use of the SIMBAD
database, operated at CDS, Strasbourg, France.  I would like to thank an
anonymous referee for helpful comments.\\

\newpage

\textbf{Glossary}\\

arcsec ($''$): A measure of angular separation.  It is one sixtieth
of an arc minute and  1/3600th of a degree.\

AU: An Astronomical Unit is the mean distance between the Earth and
the Sun.  1 AU = 1.495979 $\times$ 10$^{11}$ m.\

CCD: charge-coupled device.\

Column density: the number of atoms, \emph{e.g.} hydrogen, in a
column of area (normally 1 cm$^{-2}$) extending from the Earth to
the star.\

$E_{B-V}$ = $(B-V) - (B-V)_0$ is the colour excess (in magnitudes -
see below) due to interstellar extinction, where $B$ and $V$ are
defined at 440 and 550 nm, respectively, and $(B-V)_0$ is the
intrinsic value for the star.\

Equivalent width,
$W_\lambda=\int_{-\infty}^{\infty}\frac{I_0(\lambda)-I(\lambda)}{I_0(\lambda)}~d\lambda.$
It is usually quoted in {\AA}.\

Extinction: A term used to describe the amount of absorption and
scattering of light by gas and dust in the interstellar medium. It
is normally quoted in magnitudes.\

Interstellar reddening: For a star lying behind an interstellar
cloud, its blue light is more strongly attenuated by interstellar
material than red light.  Therefore the star appears to have a
redder colour than its instrinsic output.\

Magnitude: a logarithmic measure of the brightness of an object or
interstellar extinction such that for two stars 1 and 2, $m_1 - m_2
= -2.5 \times log_{10} [F_1/F_2]$, where $F$ is the radiation flux.\

pc = parsec: derived from \textbf{\emph{par}}allax of one
arc\textbf{\emph{sec}} and is the distance at which two objects,
separated by one AU appear to be separated by one arcsec. It is
3.085678 $\times$ 10$^{16}$ m
or $\sim$ 3.26 light years\\

\newpage

%\begin{figure}[ht]
%  \begin{center}
%   \includegraphics[angle=-90,width=17cm]{Herbig1975.ps}
%   \caption{}
%  \end{center}
%\end{figure}

\begin{figure}[ht]
\begin{center}
    \includegraphics[angle=0,width=15cm]{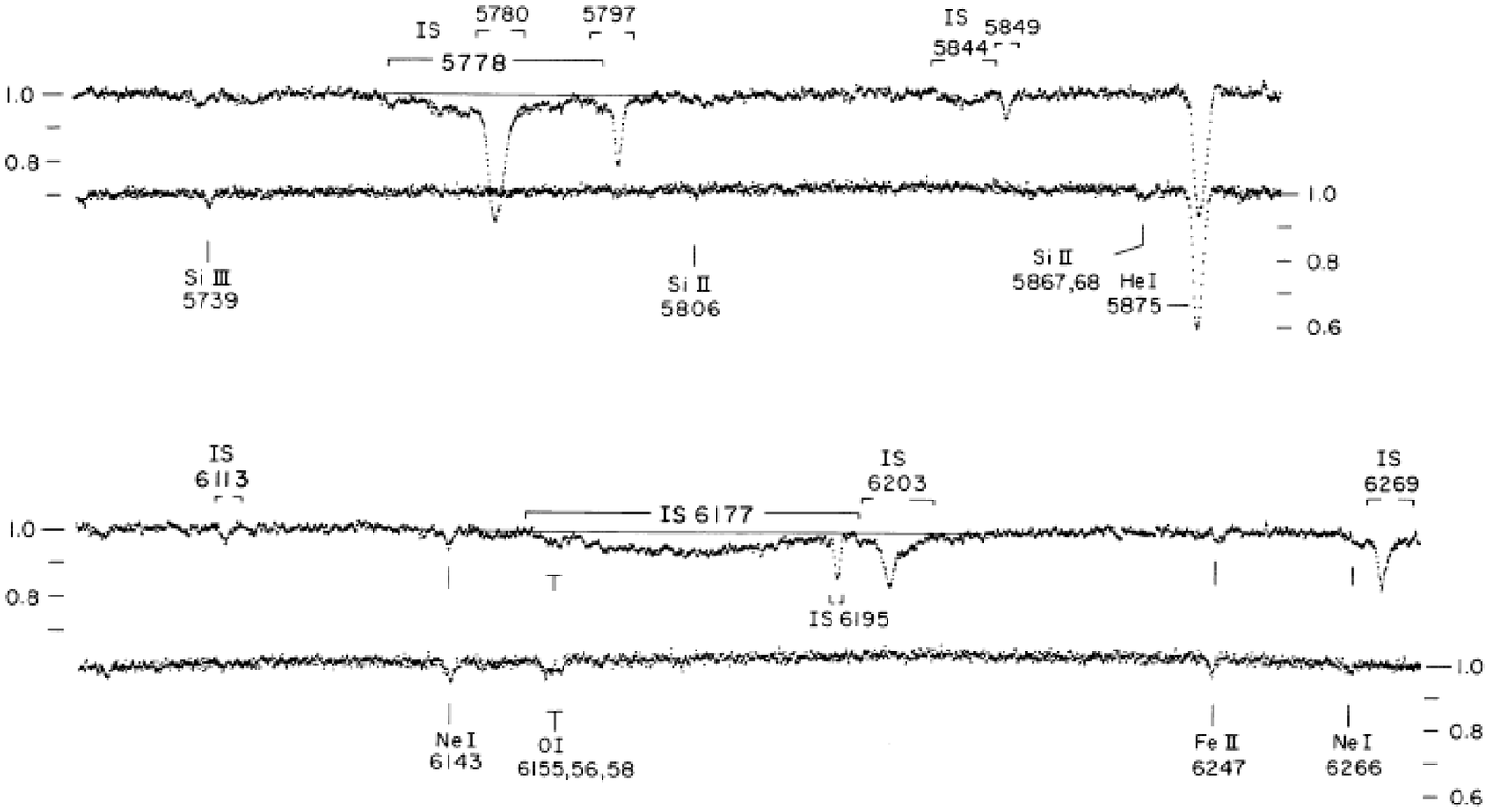}
    \caption{Comparison of spectra towards the reddened star HD 183143 (upper of each pair of traces)
    with an unreddened standard (lower).  IS indicates diffuse bands of interstellar
    origin. Note the He~\textsc{i} photospheric line that appears in
    the spectra of both stars. Reproduced with permission of the AAS
    from \cite{her75}.}
\end{center}
\end{figure}

\begin{figure}[ht]
%\begin{center}
    \includegraphics[angle=0,width=16cm]{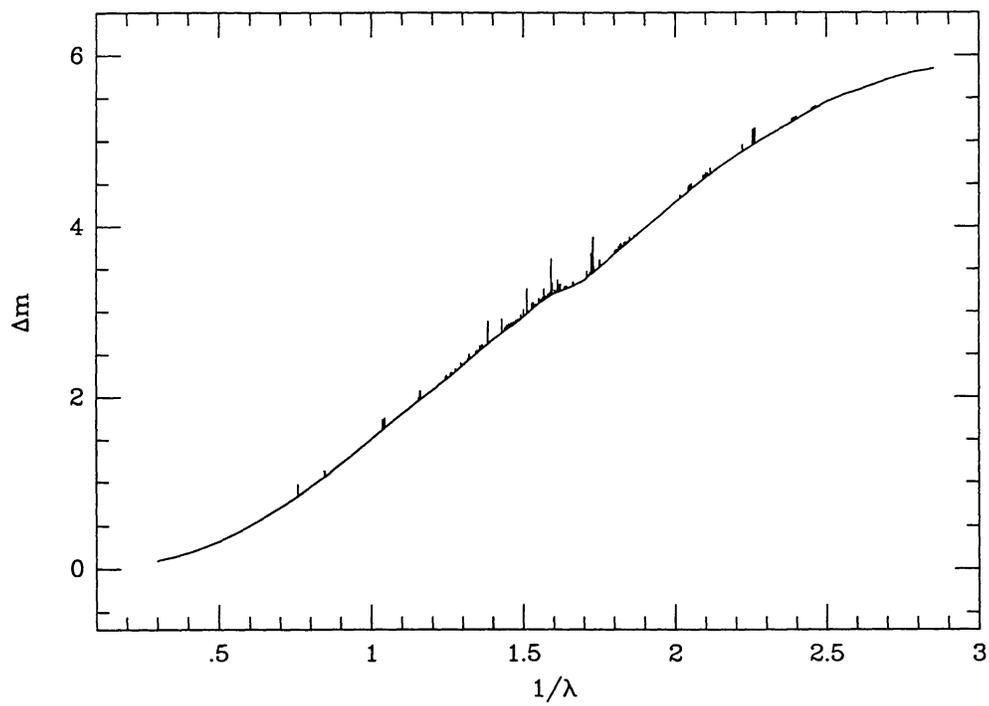}
    \caption{Diffuse interstellar bands as `fine structure' on the interstellar extinction curve of HD 183143.
    The y-axis units are magnitudes and x-axis units cm$^{-1}$ x 10$^{-4}$.
    Reproduced with permission from \cite{her95}.}
%\end{center}
\end{figure}

\begin{figure}[ht]
%\begin{center}
    \includegraphics[angle=0,width=12cm]{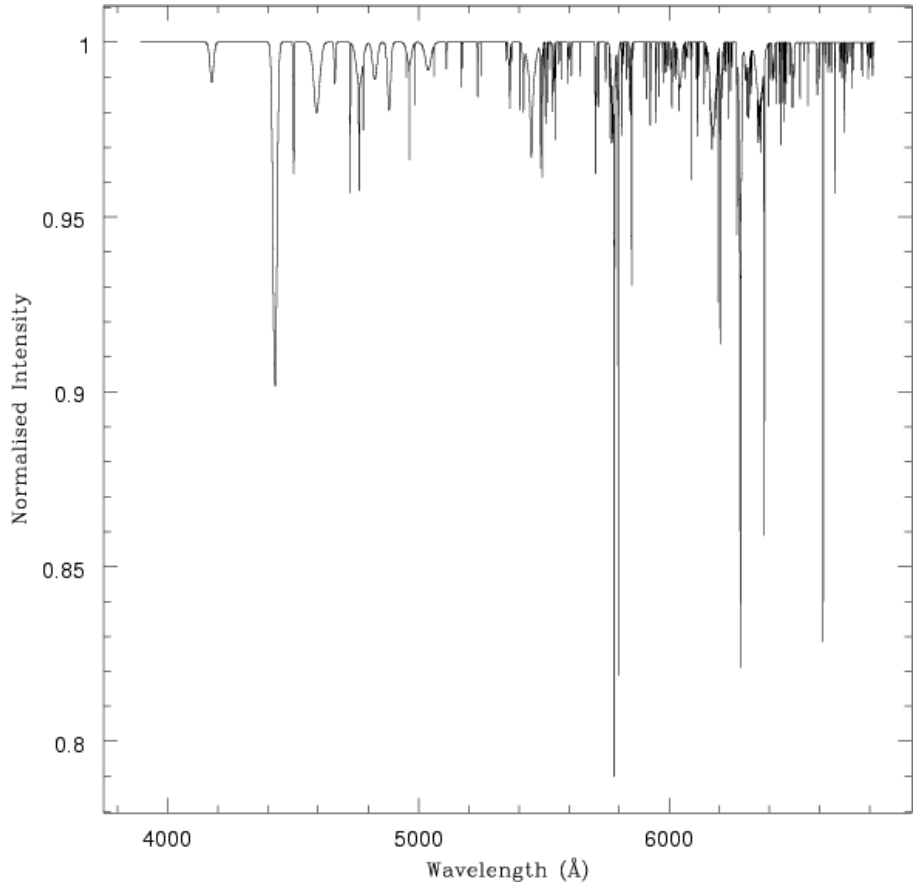}
    \caption{Representation of 226 diffuse bands observed towards BD+63$^0$~1964
    between 3906 \AA\ and 6812 \AA. Reproduced with permission from
    \cite{tua00}.}
%\end{center}
\end{figure}

\begin{figure}[ht]
% \begin{center}
    \includegraphics[angle=0,width=17cm]{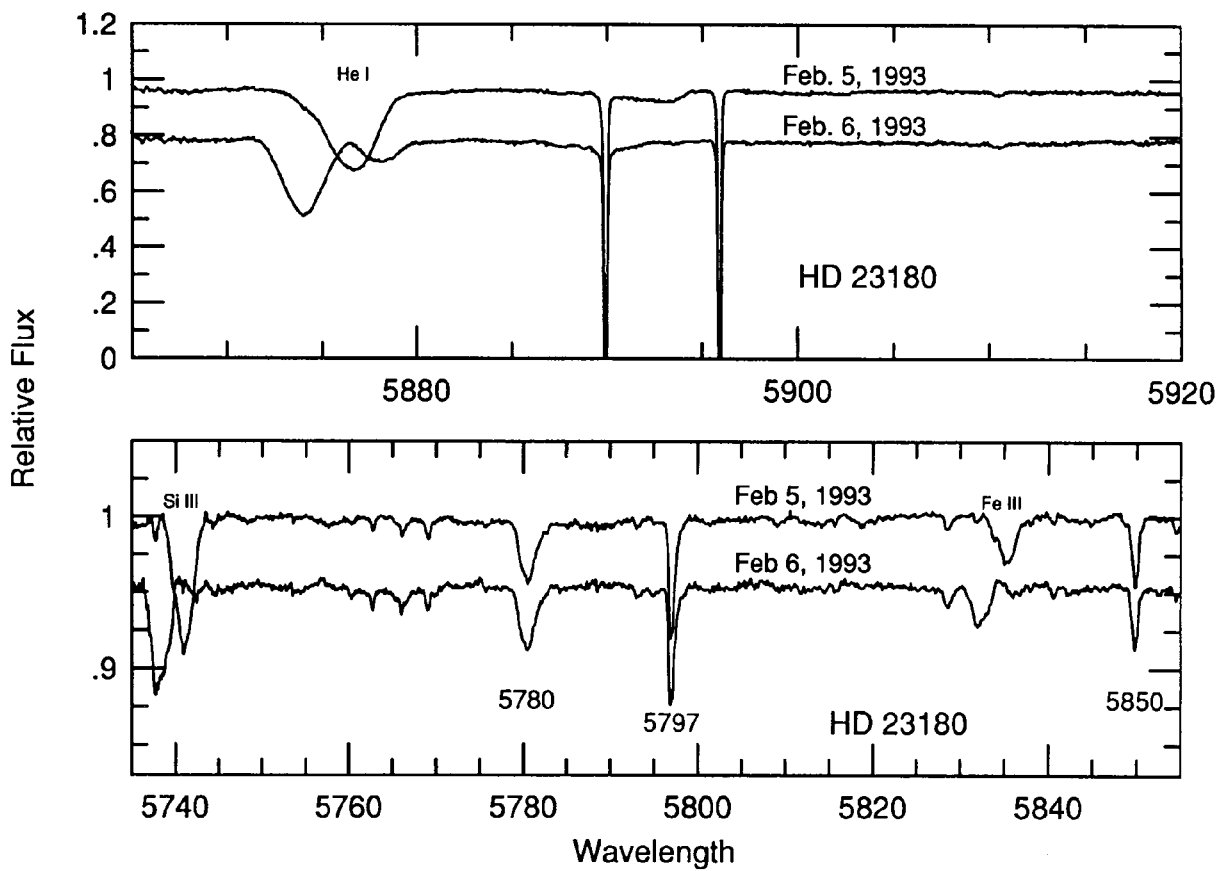}
    \caption{Spectra of the binary HD~23180 recorded on successive nights.
    The stellar He~\textsc{i}, Si~\textsc{iii} and Fe~\textsc{iii} lines are
    Doppler-shifted whereas the diffuse bands and Na~\textsc{i} lines are stationary.
    This figure originally appeared in the Publications of the Astronomical Society
    of the Pacific (J. Kre{\l}owski \& C. Sneden, 1993, PASP, 105, 1146). Copyright [1993],
    Astronomical Society of the Pacific; reproduced with permission of the Editors.}
% \end{center}
\end{figure}

\begin{figure}[ht]
% \begin{center}
    \includegraphics[angle=0,width=10cm]{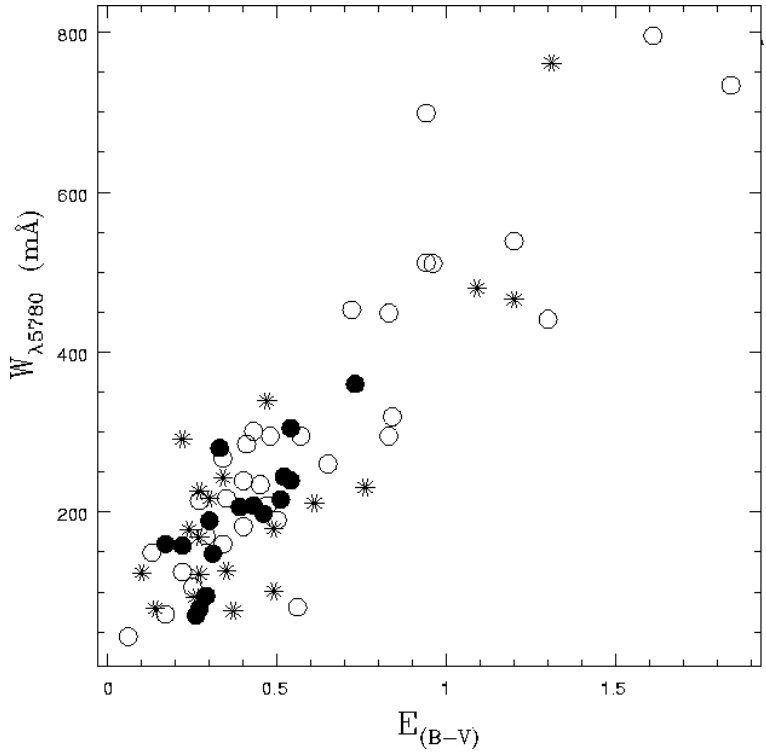}
    \caption{Plot of the equivalent width \emph{W$_{\lambda}$} of the $\lambda$5780
    diffuse band \emph{vs.} $E_{B-V}$ where the three symbols each indicate
    the average of different observations (see \cite{kre99} for details). Reproduced and adapted with permission from
    \cite{kre99}.}
    % \end{center}
\end{figure}

\begin{figure}[ht]
 \begin{center}
    \includegraphics[angle=0,width=10cm]{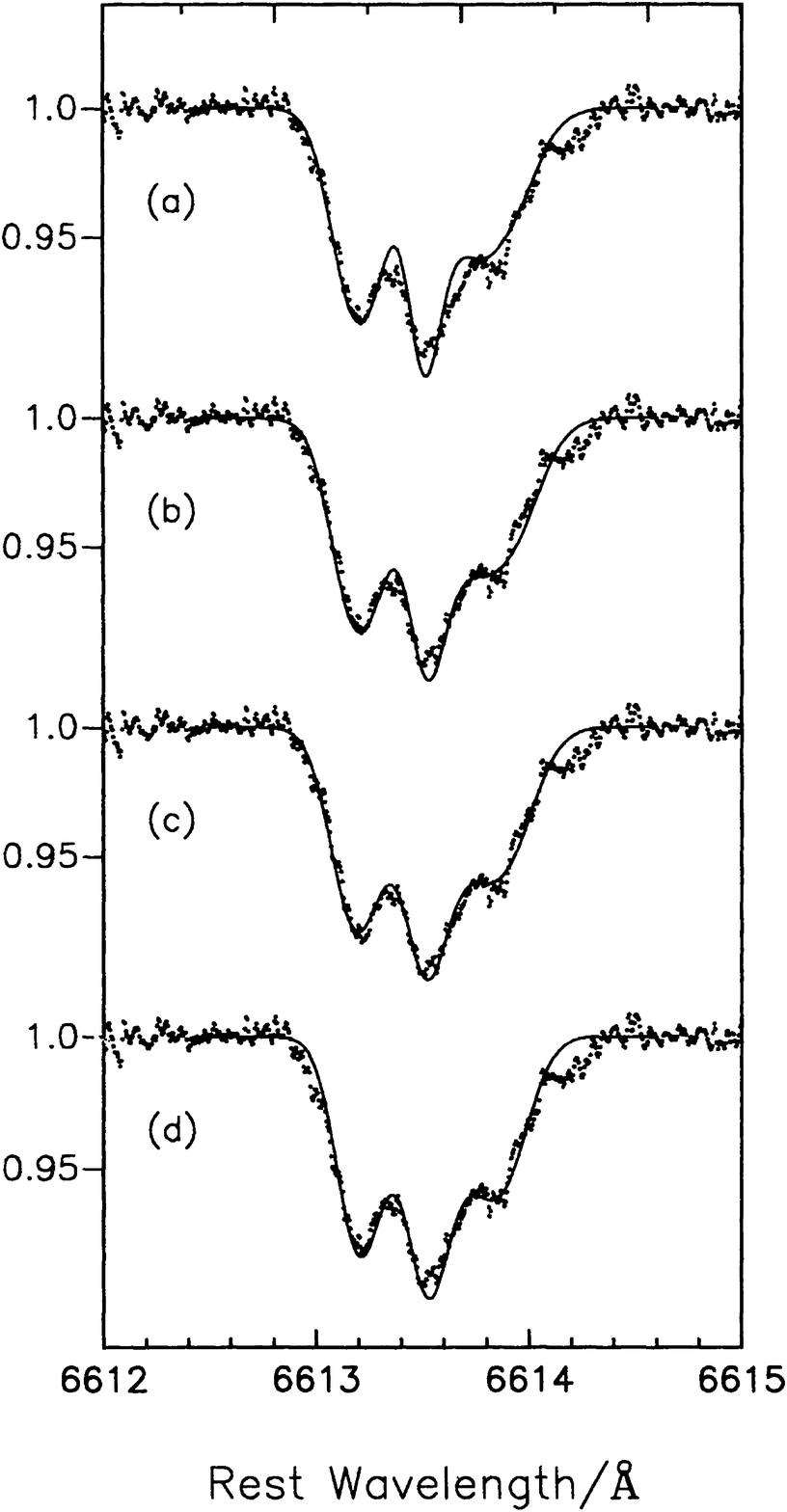}
    \caption{Ultra-high resolution spectrum of the $\lambda$6614 diffuse band showing fine structure.
    The solid lines in figures (a) to (d) are fits from different starting parameters corresponding to
    a range of molecular size/rotational temperature ($\sim$10-100\,K) combinations. See \cite{ker96} for details.
    Reproduced with permission from \cite{ker96}.}
\end{center}
\end{figure}

\begin{figure}[ht]
 \begin{center}
    \includegraphics[angle=0,width=16cm]{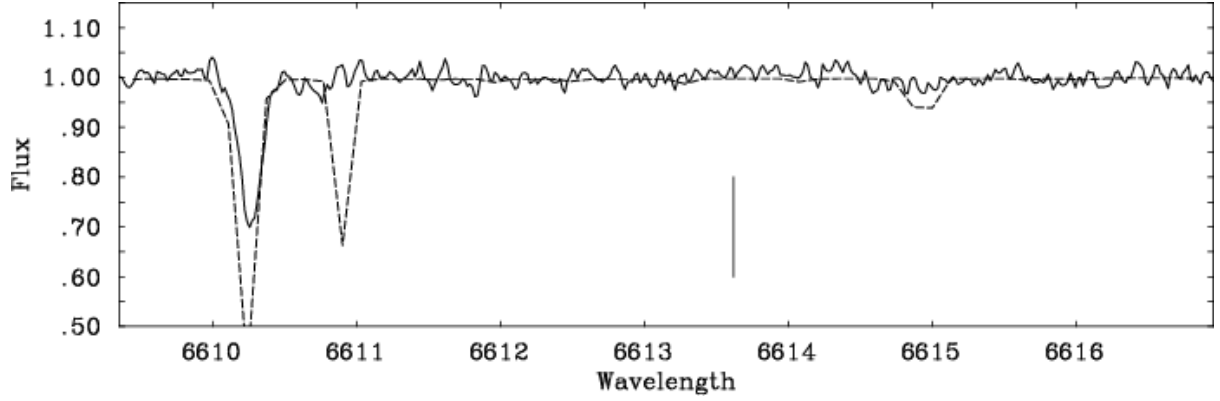}
    \caption{Spectrum taken towards Star 6 which lies behind the circumstellar shell of
    IRC+10$^o$216. No absorption at the expected position for $\lambda$6614 is found.
    The feature on the left hand side is a stellar line and the dotted line is a synthesised spectrum.
    Reproduced with permission from \cite{ken02}.}
\end{center}
\end{figure}

\begin{figure}[ht]
 \begin{center}
    \includegraphics[angle=0,width=10cm]{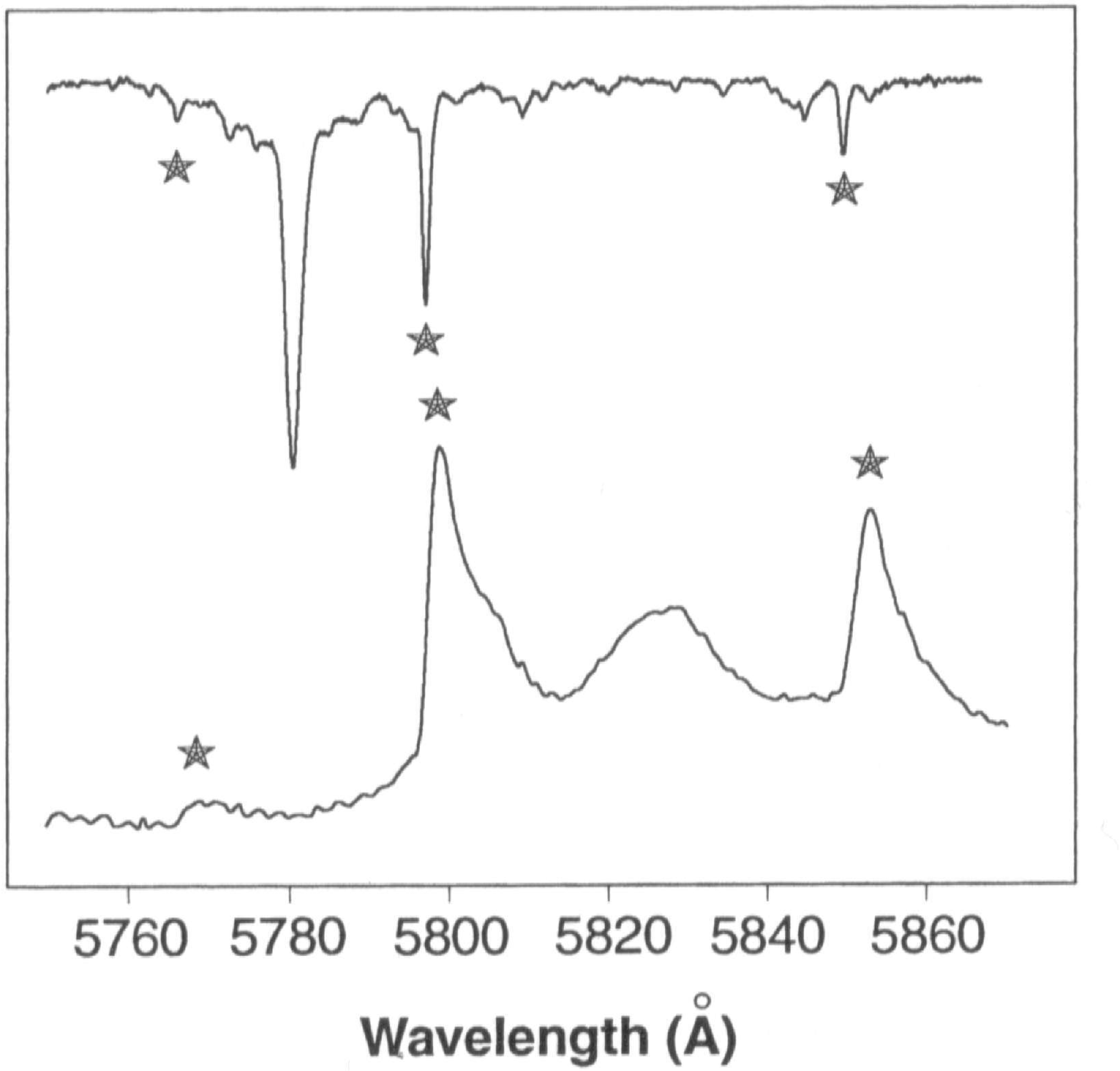}
    \caption{Comparison between the diffuse band absorption spectrum towards HD~183143 (upper)
    and the Red Rectangle emission spectrum (lower). Reproduced with permission from \cite{sar95a}.}
\end{center}
\end{figure}

\begin{figure}[ht]
% \begin{center}
    \includegraphics[angle=0,width=14cm]{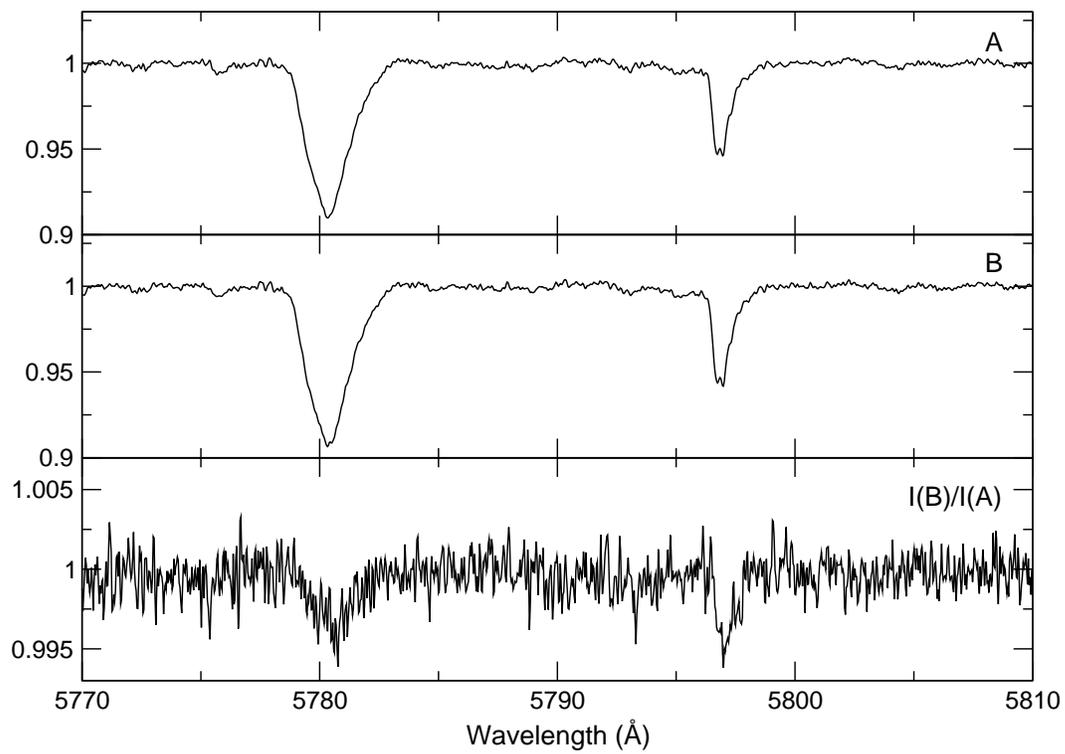}
    \caption{Upper trace: spectrum in the region of the $\lambda5780$
   and $\lambda5797$ diffuse bands towards $\rho$ Oph A; middle trace:
   spectrum towards $\rho$ Oph B; lower trace: the \emph{ratio} I$_{\lambda}$(B)/I$_{\lambda}$(A) of the spectra towards
   $\rho$ Oph A and $\rho$ Oph B. The diffuse bands are stronger towards
   B by \emph{c.} 5\%.}
%\end{center}
\end{figure}

\end{document}